\documentstyle[epsfig,prl,aps]{revtex}
\begin{document}
\draft \twocolumn[\hsize\textwidth\columnwidth\hsize\csname
@twocolumnfalse\endcsname
\title{Spherically symmetric space-time with
the regular de Sitter center}
\author{Irina Dymnikova
}
\address{Department of Mathematics and Computer Science,
         University of Warmia and Mazury,\\
Zolnierska 14, 10-561 Olsztyn, Poland; e-mail:
irina@matman.uwm.edu.pl}

\maketitle

\begin{abstract}
The requirements are formulated which lead to the existence of the
class of globally regular solutions to the minimally coupled GR
equations asymptotically de Sitter at the center. \footnote{The
extended version of the plenary talk at the V International
Friedmann Seminar on Gravitation and Cosmology, 23-30 May 2002,
Brazil. To appear in Int.J.Mod.Phys.D.} The source term for this
class, invariant under boosts in the radial direction, is
classified as spherically symmetric vacuum with variable density
and pressure $T_{\mu\nu}^{vac}$ associated with an $r-$dependent
cosmological term $\Lambda_{\mu\nu} =8\pi GT_{\mu\nu}^{vac}$,
whose asymptotic in the origin, dictated by the weak energy
condition, is the Einstein cosmological term $\Lambda g_{\mu\nu}$,
while asymptotic at infinity is de Sitter vacuum with
$\lambda<\Lambda$ or Minkowski vacuum. For this class of metrics
the mass $m$ defined by the standard ADM formula is related to
both de Sitter vacuum trapped in the origin and to breaking of
space-time symmetry. In the case of the flat asymptotic,
space-time symmetry changes smoothly from the de Sitter group at
the center to the Lorentz group at infinity through radial boosts
in between. Geometry is asymptotically de Sitter as $r\rightarrow
0$ and asymptotically Schwarzschild at large $r$. In the range of
masses $m\geq m_{crit}$, de Sitter-Schwarzschild geometry
describes  a vacuum nonsingular black hole ($\Lambda$BH), and for
$m<m_{crit}$ it describes G-lump - a vacuum selfgravitating
particlelike structure without horizons. In the case of de Sitter
asymptotic at infinity, geometry is asymptotically de Sitter as
$r\rightarrow 0$ and asymptotically Schwarzschild-de Sitter at
large $r$. $\Lambda_{\mu\nu}$ geometry describes, dependently on
parameters $m$ and $q=\sqrt{\Lambda/\lambda}$ and choice of
coordinates, a vacuum nonsingular cosmological black hole,
selfgravitating particlelike structure at the de Sitter background
$\lambda g_{\mu\nu}$, and regular cosmological models with
cosmological constant evolving smoothly from $\Lambda$ to
$\lambda$.
\end{abstract}
\pacs{PACS numbers: 04.70.Bw, 04.20.Dw}

\vskip0.2in ]

\section{ Introduction }

The well known Schwarzschild metric is the vacuum spherically
symmetric solution with a central singularity inevitable if
conditions of validity of singularity theorems are satisfied
\cite{HE}. Any other vacuum spherically symmetric solution is
reduced, by the Birkhoff theorem, to the Schwarzschild solution in
the case $T_{\mu\nu}=0$, or to the Kottler-Trefftz solution
\cite{kot} known as the Schwarzschild-de Sitter geometry \cite{GH}
in the case when a cosmological constant is included.

The non-singular modification of the Schwarzschild geometry can be
obtained by replacing a singularity with a regular core
asymptotically de Sitter as $r\rightarrow 0$. The idea goes back
to the 1966 papers of Sakharov \cite{sakharov} who considered
$p=-\rho$ as the equation of state for superhigh densities, and of
Gliner who suggested that a vacuum associated with the Einstein
cosmological term $\Lambda g_{\mu\nu}$ ($\mu$-vacuum in his
terms), could be a final state in a gravitational collapse
\cite{gliner}.

In 1991 Morgan has considered a black hole in a simple model for
quantum gravity in which quantum effects are represented by an
upper cutoff on the curvature, and obtained de Sitter-like past
and future cores replacing singularities \cite{morgan}. In 1992
Strominger demonstrated the possibility of natural, not {\it ad
hoc}, arising of de Sitter core inside a black hole in the model
of two-dimensional dilaton gravity conformally coupled to $N$
scalar fields \cite{strominger}.

Direct matching of Schwarzschild metric to de Sitter metric within
a short transitional space-like layer of the Planckian depth
\cite{markov,bernstein,farhi,shen,valera} results in metrics
typically with a jump at the junction surface.

In 1988 Poisson and Israel proposed to introduce a transitional
layer of "non-inflationary material" of uncertain depth where {\it
geometry can be self-regulatory} and describable semiclassically
down a few Planckian radii by the Einstein equations
$$
G_{\mu\nu}=-8\pi G T_{\mu\nu}           \eqno(1)
$$
with a source
term representing vacuum polarization effects \cite{werner}.

Generic properties of "noninflationary material" have been
considered in Ref.\cite{me92}. For a smooth de
Sitter-Schwarzschild transition a source term satisfies
\cite{me92}
$$
T_t^t=T_r^r;~~~T_{\theta}^{\theta}=T_{\phi}^{\phi}        \eqno(2)
$$
and the equation of state, following from $T^{\mu}_{\nu;\mu}=0$,
is
     $$p_r=-\rho;~~~p_{\perp}=-\rho-\frac{r}{2}\rho^{\prime}             \eqno(3)
     $$
The stress-energy tensor with the algebraic structure (2) has an
infinite set of comoving reference frames and is identified as
describing a spherically symmetric vacuum $T_{\mu\nu}^{vac}$,
invariant under boosts in the radial direction \cite{me92} (for
review \cite{spb,moscow,napoli}).

The exact analytical solution was found for the case of the
density profile \cite{me92}
       $$\rho(r)=\rho_0 e^{-r^3/r_0^2 r_g};
       ~~r_0^2=3/\Lambda;~~r_g=2Gm                             \eqno(4)
       $$
describing a de Sitter-Schwarzschild transition in a simple
semiclassical model for vacuum polarization in the spherically
symmetric gravitational field \cite{me96}.

In the course of Hawking evaporation a vacuum nonsingular black
hole evolves towards a self-gravitating particle-like vacuum
structure without horizons \cite{me96}, kind of gravitational
vacuum soliton called G-lump \cite{me2002}.

Model-independent analysis of the Einstein spherically symmetric
minimally coupled equations has shown \cite{me2002} which geometry
they can describe in principle if certain general requirements are
satisfied: a) regularity of metric and density at the center; b)
asymptotic flatness at infinity and finiteness of the ADM mass; c)
dominant energy condition for $T_{\mu\nu}$.

The requirements (a)-(c) define the family of asymptotically flat
solutions with the regular center which includes the class of
metrics asymptotically de Sitter as $r\rightarrow 0$ and
asymptotically Schwarzschild at large $r$. A source term connects
de Sitter vacuum in the origin with the Minkowskli vacuum at
infinity. Space-time symmetry changes smoothly from de Sitter
group at the center to the Lorentz group at infinity through the
radial boosts in between, and the standard formula for the ADM
mass relates it (generically, since a matter source can be any
from considered class) to both de Sitter vacuum trapped inside an
object and breaking of space-time symmetry \cite{me2002}.

 This class of metrics is extended to the case of non-zero
cosmological term at infinity \cite{us97} corresponding to
extension of the Einstein cosmological term $\Lambda g_{\mu\nu}$
to an $r-$dependent second rank symmetric tensor
$\Lambda_{\mu\nu}=8\pi G T_{\mu\nu}^{vac}$  connecting smoothly
two de Sitter vacua with different values of a cosmological
constant \cite{me2000}. In this approach a constant scalar
$\Lambda$ associated with a vacuum density $\Lambda=8\pi
G\rho_{vac}$, becomes a tensor component $\Lambda^t_t$ associated
explicitly with a density component of a perfect fluid tensor
whose vacuum properties follow from its symmetry and whose
variability follows from the Bianchi identities.

In this paper we review $\Lambda_{\mu\nu}$ geometry and its
application, but first we show that requirements leading to
existence of such a geometry, can be loosed to a) regularity of
density, b) finiteness of mass and c) dominant energy condition on
a stress-energy tensor. This question is addressed in Section 2.
 In Section 3 we outline geometry asymptotically flat at infinity, de Sitter-Schwarzschild
 geometry. Section 4 is devoted to geometry asymptotically de Sitter at both
 origin and infinity. Section 5 contains summary and discussion.

\section{$\Lambda_{\mu\nu}$ geometry}

\subsection{Basic equations }

A static spherically symmetric line element can be written in the
form \cite{tolman}
     $$
     ds^2 = e^{\mu(r)}dt^2 - e^{\nu(r)} dr^2 - r^2 d\Omega^2
                                                                       \eqno(5)
     $$
where $d\Omega^2$ is the metric of a unit 2-sphere. The metric
coefficients satisfy the Einstein equations (1) which reduce to
    $$
    \kappa T_t^t = \kappa\rho(r)= e^{-\nu}\biggl(\frac{{\nu}^{\prime}}{r}
    -\frac{1}{r^2}\biggr)+\frac{1}{r^2}
                                                                            \eqno(6)
    $$
    $$\kappa T_r^r =-\kappa p_r(r)= -e^{-\nu} \biggl(\frac{{\mu}^{\prime}}{r}
     +\frac{1}{r^2}\biggr)+\frac{1}{r^2}
                                                                              \eqno(7)
    $$
    $$\kappa T_{\theta}^{\theta}=\kappa T_{\phi}^{\phi}=-\kappa p_{\perp}(r)=$$
    $$-e^{-\nu}\biggl(\frac{{{\mu}^{\prime\prime}}}{2}
    +\frac{{{\mu}^{\prime}}^2}{4}
    +\frac{({{\mu}^{\prime}-{\nu}^{\prime}})}{2r}-\frac{{\mu}^{\prime}
    {\nu}^{\prime}}{4}\biggr)
                                                                                \eqno(8)
    $$
Here $\kappa = 8\pi G$  (we adopted $c=1$ for simplicity),
$\rho(r)=T^t_t$ is the energy density, $p_r(r)=-T^r_r$ is the
radial pressure, and
$p_{\perp}(r)=-T_{\theta}^{\theta}=-T_{\phi}^{\phi}$ is the
tangential pressure for anisotropic perfect fluid \cite{tolman}.
The prime denotes differentiation  with respect to $r$.
 Integration of Eq.(6) gives
    $$e^{-\nu(r)}=g(r)=1-\frac{2GM(r)}{r};~~M(r)
    =4\pi\int_0^r{\rho(x)x^2dx}
                                                                              \eqno(9)
    $$
whose asymptotic for large $r$ is $e^{-\nu}=1-{2Gm}/{r}$, with the
mass parameter
   $$
   m=4\pi\int_0^{\infty}{\rho(r) r^2 dr}
                                                                               \eqno(10)
   $$
Equations (6)-(8) give the Oppenheimer equation \cite{oppi}
$$
     \kappa(T^t_t-T^r_r)=\kappa(p_r+\rho)=\frac{e^{-\nu}}{r}
     (\nu^{\prime}+\mu^{\prime})
                                                                                 \eqno(11)
     $$
and hydrodynamic equation which generalizes the
Tolman-Oppenheimer-Volkoff equation \cite{wald} to the case of
different principal pressures \cite{apj,me2002}
     $$
     p_{\perp}=p_r+\frac{r}{2}p_r^{\prime}+(\rho+p_r)\frac{G M(r)
     +4\pi G r^3 p_r}{2(r-2G M(r))}
                                                                                   \eqno(12)
     $$

To investigate the system we impose the following requirements:

a)Regularity of a density $\rho(r)$

b) Finiteness of the mass parameter $m$

c) Dominant Energy Condition (DEC) on $T_{\mu\nu}$

 The dominant energy condition, $T^{00}\geq|T^{ab}|$,
$a,b=1,2,3$, holds if and only if \cite{HE}
     $$\rho\geq0;~~~~-\rho\leq p_k\leq \rho;~~~~k=1,2,3
                                                                                    \eqno(13)
     $$
and implies that the energy density as measured by any local
observer moving along a time-like curve, is non-negative, and each
principal pressure does not exceed the energy density. The Weak
Energy Condition (WEC) contained in the DEC, reads
$T_{\mu\nu}\xi^{\mu}\xi^{\nu}\geq 0$ for any time-like vector
$\xi^{\mu}$, and holds if and only if
$$
\rho\geq 0,~~~ \rho+p_k\geq 0, ~~~  k=1,2,3
                                                                              \eqno(14)
$$

The requirements a)-c) imposed on a system (6)-(8), enforce the
following behavior. Finiteness of a mass (10) leads to $\nu(r)=0$
as $r\rightarrow\infty$, and requires a density profile $\rho(r)$
to vanish at infinity quicker than $r^{-3}$. DEC requires $p_k$ to
vanish equally fast or faster as $r\rightarrow\infty$. By the
Oppenheimer equation (11), $\mu^{\prime}=0$ and $\mu=$const at
infinity. Rescaling the time coordinate leads to the standard
boundary condition $\mu\rightarrow 0$ as $r\rightarrow \infty$,
asymptotic flatness needed to identify (10) as the ADM mass
\cite{wald}.

 Regularity of density $\rho(r=0)<\infty$, requires the mass
 function $M(r)$ to vanish as $r^3$ when $r\rightarrow 0$,
 as a result $\nu(r)\rightarrow 0$ as $r\rightarrow 0$.
 It leads also, by DEC,
to regularity of pressures, then $p_r+\rho < \infty$ leads to
$\nu^{\prime}+\mu^{\prime}=0$ and $\nu+\mu=\mu(0)$ at $r=0$ with
$\mu(0)$ playing the role of the family parameter.

The weak energy condition defines, by the Oppenheimer equation
(11), the sign of the sum $\mu^{\prime}+\nu^{\prime}$. It demands
$\mu^{\prime}+\nu^{\prime}\geq 0$ everywhere \cite{me2002}(in the
regions inside the horizons, the radial coordinate $r$ is
time-like and $T_t^t$ represents a tension, $p_r=-T_t^t$, along
the axes of the space-like 3-cylinders of constant time $r$=const
\cite{werner}, then $T_t^t-T_r^r=-(p_r+\rho)$). As a result the
function $\mu+\nu$ is growing from $\mu=\mu(0)$ at $r=0$ to
$\mu=0$ at $r\rightarrow\infty$, which gives $\mu(0)\leq 0$
\cite{me2002}.

The example of solution from this family is boson stars
\cite{boson} (for review \cite{Mielke}) which are regular
configurations without horizons generated by a self-gravitating
massive scalar field whose stress-energy tensor is essentially
anisotropic, $p_r\neq p_{\perp}$.

The range of family parameter $\mu(0)$ includes the value
$\mu(0)=0$. In this case the function $\nu(r)+\mu(r)$ is zero at
$r=0$ and at $r\rightarrow\infty$, its derivative is non-negative,
it follows that  $\nu(r)=-\mu(r)$ everywhere.

For this class of metrics behavior at $r\rightarrow 0$ is dictated
by the WEC. It is easily to prove \cite{me2002} that the function
$\mu(r)+\nu(r)$ cannot have extremum at $r=0$, therefore
$\mu^{\prime\prime}+\nu^{\prime\prime}=0$ at $r=0$. It leads to
$p_r+\rho=0$ at $r=0$, and  in the limit $r\rightarrow 0$ Eq.(12)
gives $p_{\perp}=-\rho-\frac{r}{2}\rho^{\prime}$. The DEC and
regularity of $\rho$ requires $p_k+\rho<\infty$ and thus
$|\rho^{\prime}|<\infty$. The equation of state near the center
becomes $p=-\rho$ \cite{me2002}, which gives de Sitter asymptotic
as $r\rightarrow 0$
$$
 ds^2=\biggl(1-\frac{r_2}{r_0^2}\biggr)dt^2
     -\frac{dr^2}{\biggl(1-\frac{r_2}{r_0^2}\biggr)}-r^2d\Omega^2
                                                                         \eqno(15a)
      $$
$$
T_{\mu\nu}=\rho_0 g_{\mu\nu};~~~~~~ \rho_0 =\kappa^{-1}\Lambda; ~~
~~ ~~ r_0^2=\frac{3}{\Lambda}
                                                                         \eqno(15b)
$$
where $\rho_0=\rho(r=0)$ and $\Lambda$ is the cosmological
constant which appeared at the origin although was not present in
the basic equations.

The weak energy condition $p_{\perp}+\rho\geq 0$ gives
$\rho^{\prime}\leq 0$, and thus demands monotonic decreasing of a
density profile. This defines the form of  the metric function
$g(r)$. In the region $0<r<\infty$ it has only minimum and the
geometry can have not more than two horizons \cite{me2002}.

Requirements a)-c) lead thus to the existence of the class of
metrics
$$
ds^2=g(r)dt^2-\frac{dr^2}{g(r)}-r^2 d\Omega^2
                                                                       \eqno(16a)
$$
$$
g(r)=1-\frac{R_g(r)}{r};  ~~~ R_g(r)=2G M(r); ~~~ M(r)
    =4\pi\int_0^r{\rho(x)x^2dx}
                                                                       \eqno(16b)
$$
It is asymptotically de Sitter as $r\rightarrow 0$, asymptotically
Schwarzschild at large $r$
$$
  ds^2=\biggl(1-\frac{r_g}{r}\biggr)-
     \frac{dr^2}{\biggl(1-\frac{r_g}{r}\biggr)}
     -r^2d\Omega^2; ~~~r_g=2Gm
                                                                    \eqno(17)
     $$
and has at most two horizons: a black hole horizon and an internal
Cauchy horizon.

This class of metrics is extended to the case of nonzero
background cosmological constant $\lambda$ by introducing
$T_t^t(r)=\rho(r)+(8\pi G)^{-1}\lambda$. The metric function
\cite{us97}
$$
g(r)=1-\frac{2GM(r)}{r} -\frac{\lambda r^2}{3}
                                                                                \eqno(18)
$$
is asymptotically   de Sitter at both origin and infinity, with
$\lambda$ as $r\rightarrow \infty$ and with $\Lambda +\lambda$ as
$r\rightarrow 0$. The source term evolves from $(\Lambda+\lambda)
g_{\mu\nu}$ to $\lambda g_{\mu\nu}$.

\subsection{Cosmological term}

For the class of metrics (16) a source term has the algebraic
structure (2) \cite{me92}. It is invariant under rotations in the
$(r,t)$ plane, has an infinite set of comoving reference frames
and is identified as describing a spherically symmetric vacuum
$T_{\mu\nu}^{vac}$, invariant under boosts in the radial direction
\cite{me92}.

For considered class of metrics it must be asymptotically de
Sitter as $r\rightarrow 0$. It connects de Sitter vacuum
$T_{\mu\nu}=\rho_0 g_{\mu\nu}$ in the origin with the Minkowski
vacuum $T_{\mu\nu}=0$ at infinity or two de Sitter vacua with
different values of cosmological constant
$$
\kappa^{-1}\Lambda g_{\mu\nu} ~~ \leftarrow~~~
T_{\mu\nu}^{vac}~~~\rightarrow  ~~ \kappa^{-1}\lambda g_{\mu\nu}
                                                               \eqno(19)
$$
In the paper on $\Lambda$-variability, Overduin and Cooperstock
distinguished two basic approaches to $\Lambda g_{\mu\nu}$
existing in the literature \cite{overduin}. In the first approach
$\Lambda g_{\mu\nu}$ is shifted onto the right-hand side of the
Einstein equations  and treated as a dynamical part of the matter
content. This approach, characterized by Overduin and Cooperstock
as connected to dialectic materialism of the Soviet physics
school, goes back to Gliner who interpreted ${\kappa}^{-1} \Lambda
g_{\mu\nu}$ as vacuum stress-energy tensor \cite{gliner}, and to
Zel'dovich who connected $\Lambda$ with the gravitational
interaction of virtual particles \cite{zeld}. The second
(idealistic) approach prefers to keep $\Lambda$ on the left-hand
side of Eqs.(1) as geometrical entity and treat it as a constant
of nature.

 In any case direct association of a
cosmological term $\Lambda g_{\mu\nu}$ with the vacuum stress
tensor $\rho_{vac} g_{\mu\nu}; ~~ \rho_{vac}={\kappa}^{-1}\Lambda$
seems widely accepted today \cite{dewitt}.

Here we started from the Einstein equation (1) and found the case
when we have on the right-hand side a stress-energy tensor
describing a spherically symmetric anisotropic vacuum with
variable density and pressures. Nothing would prevent from
shifting $T_{\mu\nu}^{vac}$ to the left-hand side of Eqs.(1) and
treating it as evolving geometrical entity.\footnote{The Einstein
equations (1) can be written in the four-indices form
$G_{\alpha\beta\gamma\delta}=-8\pi G T_{\alpha\beta\gamma\delta}$
as the equivalence relations which put the matter and geometry in
direct algebraic correspondence \cite{us83}.} A variable
cosmological term \cite{me2000}
$$
\Lambda_{\mu\nu}=\kappa T_{\mu\nu}^{vac}
                                                                    \eqno(20)
$$
satisfies the
 Einstein equation (1) which in this case reads
$$
G_{\mu}^{\nu}+\Lambda_{\mu}^{\nu}=0
                                              \eqno(21)
$$
and the Bianchi identities determine its evolution
$$
\Lambda_{\nu;\mu}^{\mu}=0
                                                \eqno(22)
$$
As a result $\Lambda_{\mu\nu}$ satisfies the equation of state (3)
with \cite{me2000}
$$
\kappa\rho^{\Lambda}(r)=\Lambda_t^t;~ ~\kappa
p^{\Lambda}_r(r)=-\Lambda^r_r;  ~~ \kappa
p^{\Lambda}_{\perp}(r)=-\Lambda_{\theta}^{\theta}
=-\Lambda^{\phi}_{\phi}                          \eqno(23)
$$
A cosmological tensor $\Lambda_{\mu\nu}$ represents the extension
of the algebraic structure of the Einstein cosmological term
$\Lambda g_{\mu\nu}$  to an $r$-dependent cosmological term
$\Lambda_{\mu\nu}$ which includes $\Lambda g_{\mu\nu}$ as the
particular case $\Lambda = \kappa\rho_{vac}$=const when the full
symmetry is restored, and remains $\Lambda g_{\mu\nu}$ as proper
asymptotics of $\Lambda_{\mu\nu}$ at both regular center and
infinity. In such extension the cosmological constant - scalar
associated with the vacuum density - becomes a tensor component
$\Lambda_t^t$ associated {\it explicite} with the density
component of vacuum stress tensor $T_{\mu\nu}^{vac}$ defined by
its symmetry.

\section{ De Sitter-Schwarzschild geometry }

De Sitter-Schwarzschild geometry \cite{me96}(for review
\cite{spb,moscow,napoli}) is asymptotically de Sitter as
$r\rightarrow 0$ and asymptotically Schwarzschild as
$r\rightarrow\infty$.

For the case of density profile (4) the metric and mass functions
are \cite{me92}
$$
g(r)=1-\frac{r_g}{r}\biggl(1-e^{-r^3/r_0^2 r_g}\biggr); ~~
M(r)=m\biggl(1-e^{-r^3/r_0^2 r_g}\biggr)
                                                       \eqno(24)
$$
Two horizons, a black hole event horizon $r_{+}$ and an internal
Cauchy horizon $r_{-}$, are shown in Fig.1 together with two
characteristic surfaces of the geometry: a zero-gravity surface
$r=r_c$ beyond which the strong energy condition of singularities
theorems \cite{HE} is violated (zero-gravity surface is defined by
$2\rho+r\rho^{\prime}=0$\cite{me96}), and a zero-curvature surface
$r=r_s$ at which scalar curvature $R(r)$ vanishes. For the case
(24) $r_s$ is given by \cite{me96}
$$r_s=\biggl(\frac{4}{3}r_0^2 r_g\biggr)^{1/3}
=\biggl(\frac{m}{\pi\rho_0}\biggr)^{1/3}
                                                                          \eqno(25)
$$
and confines about 3/4 of the mass $m$.
\begin{figure}
\vspace{-8.0mm}
\begin{center}
\epsfig{file=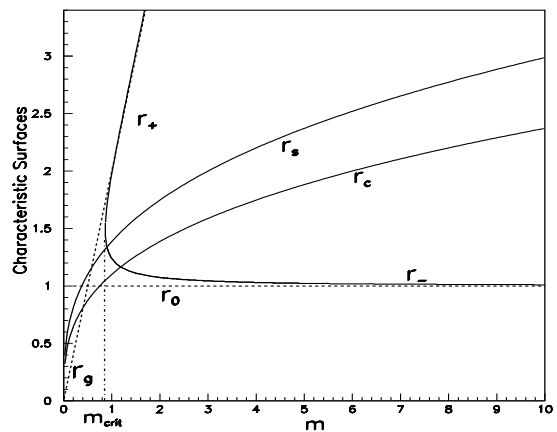,width=8.0cm,height=6.5cm}
\end{center}
\caption{Characteristic surfaces of de Sitter-Schwarzschild
geometry } \label{fig.1}
\end{figure}
Horizons come together at the value of a mass parameter
$m_{crit}$, which puts a lower limit on a black hole mass
\cite{me96}. For the case of a density profile (4)
 the critical mass is \cite{me92,me96}
$$
m_{crit}\simeq{0.3 m_{Pl}\sqrt{\rho_{Pl}/\rho_0}}
                                                                        \eqno(26)
$$
For $m<m_{crit}$ geometry describes a self-gravitating
particle-like structure without horizons. De Sitter-Schwarzschild
configurations are shown in Fig.2.
\begin{figure}
\vspace{-8.0mm}
\begin{center}
\epsfig{file=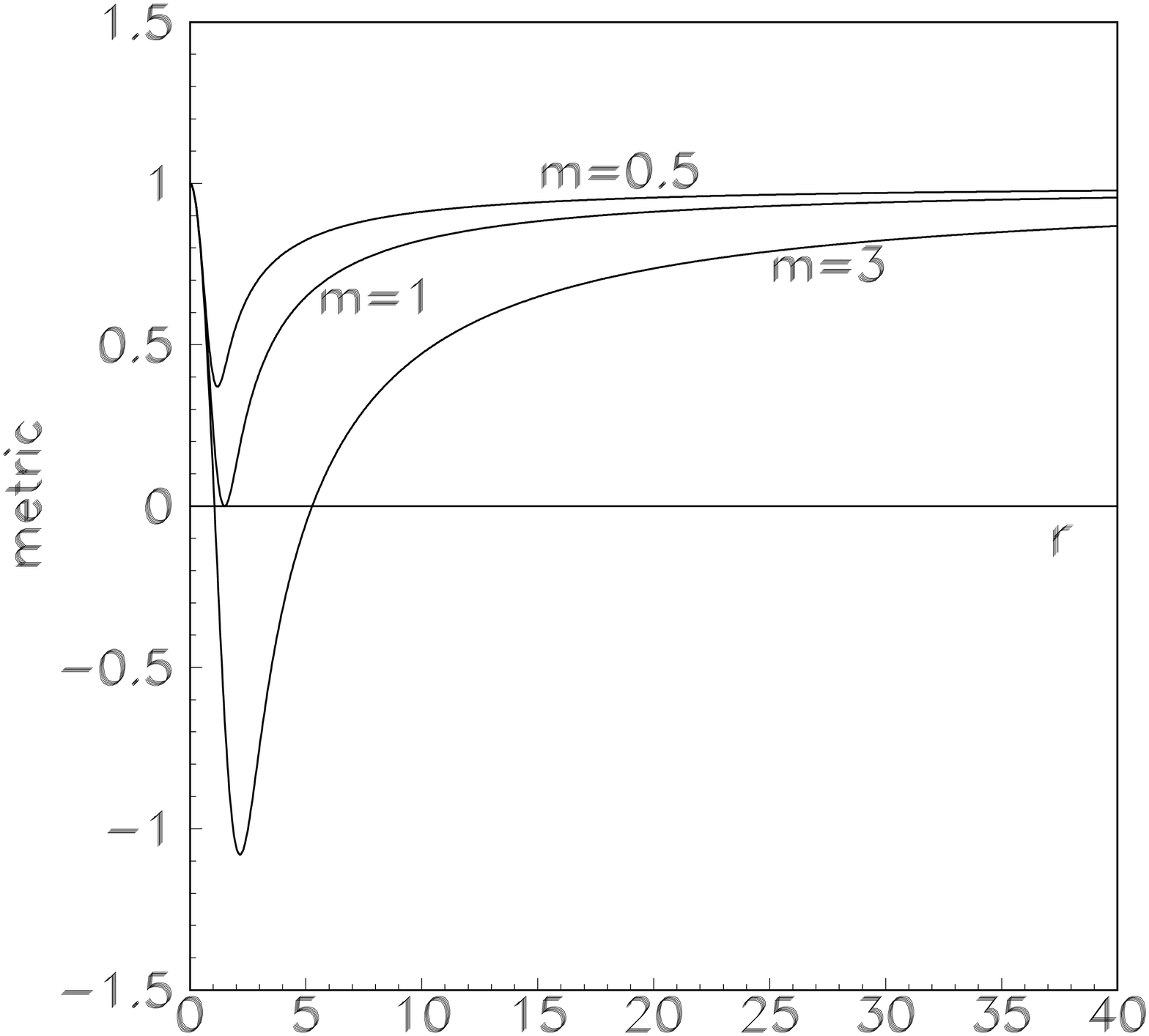,width=8.0cm,height=5.5cm}
\end{center}
\caption{ The metric $g(r)$ for de Sitter-Schwarzschild
configurations. The mass $m$ is normalized to $m_{crit}$. }
\label{fig.2}
\end{figure}
For $m\geq m_{crit}$ de Sitter-Schwarzschild geometry describes
the vacuum nonsingular black hole, and global structure of
space-time shown in Fig.3,  contains an infinite sequence of black
and white holes whose future and past singularities are replaced
with regular cores asymptotically de Sitter as $r\rightarrow 0$
\cite{me96}. It resembles the Reissner-Nordstr\"om case, but
differs essentially at the surfaces $r=0$ which are regular
surfaces of de Sitter-Schwarzschild geometry.

 Vacuum nonsingular
black and white holes can be called $\Lambda$BH,  $\Lambda$WH,
since de Sitter vacuum ($\Lambda g_{\mu\nu}$) appears instead of a
singularity at approaching each regular surface $r=0$.

\subsection{Vacuum nonsingular white hole}

Replacing a Schwarzschild singularity with the regular core
transforms the space-like singular surfaces $r=0$ of Schwarzschild
geometry into the time-like regular surfaces $r=0$ in the future
of a $\Lambda$BH and in the past of a $\Lambda$WH. In a sense this
rehabilitates a white hole whose existence in a singular version
has been forbidden by the cosmic censorship since a singularity
open into the future of a universe $U$ breaks the predictability
in $U$ \cite{penrose}. In the case of $\Lambda$WH predictability
is only restricted by the existence of the Cauchy horizon in de
Sitter-Schwarzschild geometry \cite{us2001}.

The regular core in the past of a $\Lambda$WH models an early
evolution of an expanding universe. The expansion starts from a
regular surface $r=0$ with a nonsingular non-simultaneous de
Sitter bang followed by a Kasner-like stage of anisotropic
expansion at which most of mass is produced \cite{us2001}.
\begin{figure}
\vspace{-8.0mm}
\begin{center}
\epsfig{file=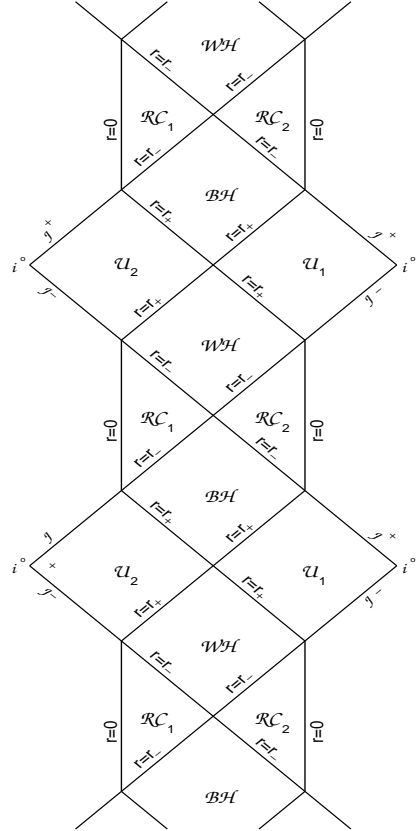,width=8.0cm,height=12.5cm}
\end{center}
\caption{Global structure of de Sitter-Schwarzschild space-time.}
\label{fig.3}
\end{figure}

\subsection{Baby universes inside a $\Lambda$BH }

The idea of a  universe inside a black hole has been suggested by
Farhi and Guth in 1987 \cite{farhi} as the idea of creation of a
universe in the laboratory starting from a false vacuum bubble in
the Minkowski space. They studied an expanding spherical de Sitter
bubble separated by a thin wall from the outside region of the
Schwarzschild geometry. The case of direct de Sitter-Schwarzschild
matching  corresponds to arising of a closed or semiclosed world
inside a black hole \cite{valera}.

In general case of a distributed density profile, situation is
different since the global structure of space-time is essentially
different. In de Sitter-Schwarzschild geometry, closed or
semiclosed world can arise in any of the regular cores near $r=0$
in the future of a $\Lambda$BH. An infinite number of such cores
inside a $\Lambda$BH enhances  the probability of arising of
closed or semiclosed world inside a $\Lambda$BH \cite{us2001}.

Global structure of space-time in case of a $\Lambda$BH leads  to
additional  possibilities related to instability of the de Sitter
vacuum near the surface $r=0$. Instability of de Sitter vacuum
with respect to quantum birth of a universe is well studied
\cite{kolb}. The essential feature is possibility of multiple
birth of causally disconnected universes from the de Sitter vacuum
noted first in 1975 in the Ref \cite{us75}. The global structure
of space-time was considered in 1982 by Gott III for the case of
creation of an open FRW universe \cite{gott}. In the context of
the minisuperspace model quantum birth of an open or flat universe
is possible when an initial quantum fluctuation contains an
admixture of radiation and strings or some other quintessence with
the equation of state $p=-\rho/3$ which mimics a curvature term.
In the presence of radiation quantum tunnelling occurs from a
discrete energy level with a nonzero quantized temperature
\cite{usbest}. An infinite number of regular cores $r=0$ inside a
$\Lambda$BH enhances essentially the probability of quantum birth
of  baby universes inside it as a result of quantum instability of
de Sitter vacuum \cite{us2001}.

\subsection{Vacuum nonsingular black hole}

A $\Lambda$BH emits Hawking radiation from both horizons with the
Gibbons-Hawking temperature \cite{GH} which for $\Lambda$BH with
two horizons is given by \cite{me96}
$$
k T=\frac{\hbar c}{4\pi} \biggl[\frac{R_g(r_h)}{r_h^2}-
\frac{R_g^{\prime}(r_h)}{r_h}\biggr]; ~~~ r_h=r_{+},r_{-}
                                                                          \eqno(27)
$$
The form of the emperature-mass diagram is generic for de
Sitter-Schwarzschild geometry. The temperature on the BH horizon
drops to zero at $m=m_{crit}$, while the Schwarzschild asymptotic
requires $T_{+}\rightarrow 0$ as $m\rightarrow\infty$. The
temperature-mass curve has thus a maximum between $m_{crit}$ and
$m\rightarrow\infty$. In a maximum the specific heat is broken and
changes its sign testifying to a second-order phase transition in
the course of Hawking evaporation and suggesting symmetry
restoration to the de Sitter group in the origin \cite{me97}.

For particular form of the density profile (4) the temperature is
given by \cite{me96}
$$
T_h=\frac{\hbar c}{4\pi k
r_0}\biggl[\frac{r_0}{r_h}-\frac{3r_h}{r_0}
\biggl(1-\frac{r_h}{r_g}\biggr)\biggr]
                                                                            \eqno(28)
$$
The mass at the maximum and the temperature of the phase
transition are \cite{me96}
$$
m_{tr}\simeq{0.38 m_{Pl}\sqrt{\rho_{Pl}/\rho_0}};  ~~
T_{tr}\simeq{0.2 m_{Pl}\sqrt{\rho_{Pl}/\rho_0}}
                                                                          \eqno(29)
$$

\subsection{G-lump}

For masses $m<m_{crit}$ de Sitter-Schwarzschild geometry describes
a self-gravitating particle-like vacuum structure, globally
regular and globally neutral.
 It resembles
Coleman's lumps - non-singular, non-dissipative solutions of
finite energy, holding themselves together by their own
self-interaction \cite{lump}. G-lump holds itself together by
gravity due to balance between gravitational attraction outside
and gravitational repulsion inside of zero-gravity surface
$r=r_c$. For the case of density profile (4) it is  perfectly
localized (see Fig.4) \cite{me2002}.
\begin{figure}
\vspace{-8.0mm}
\begin{center}
\epsfig{file=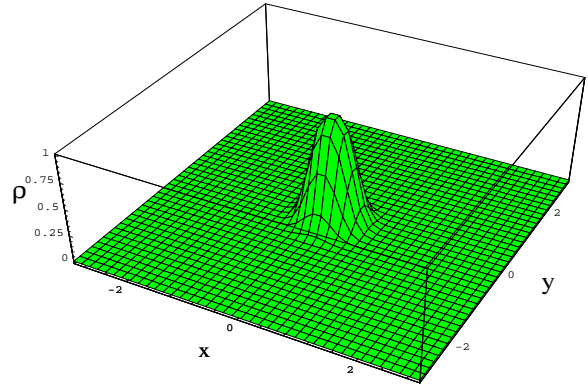,width=8.0cm,height=5.5cm}
\end{center}
\caption{ G-lump in the case $r_g=0.1r_0$ ($m\simeq{0.06
m_{crit}})$. }
\label{fig.4}
\end{figure}
Since de Sitter vacuum is trapped within a G-lump, it can be
modelled  by a spherical bubble with monotonically decreasing
density. Its geometry is described by the metric \cite{me2002}
$$
 ds^2=d\tau^2-\frac{2GM(r(R,\tau))}{r(R,\tau)}-r^2(R,\tau)d\Omega^2
                                                                             \eqno(30)
$$
The equation of motion \cite{us2001} has the first integral
\cite{me2002}
$$
\dot{r}^2 - \frac{2GM(r)}{r}=f(R)
                                                                              \eqno(31)
$$
which resembles the equation of a particle in the potential
$V(r)=-\frac{GM(r)}{r}$, with the constant of integration $f(R)$
playing the role of the total energy $f=2E$.

A spherical bubble can be described by the minisuperspace model
with a single degree of freedom \cite{vil}. Zero-point vacuum
energy for G-lump, which clearly represents an elementary
spherically symmetric excitation of a vacuum defined
macroscopically by its symmetry (2), is evaluated as its minimal
quantized energy.

By the standard procedure of quantization the equation (26)
transforms into the Wheeler-DeWitt equation in the minisuperspace
\cite{vil}
$$
\frac{\hbar^2}{2m_{Pl}}\frac{d^2\psi}{dr^2}-(V(r)-E)\psi=0
                                                                               \eqno(32)
$$
Near the minimum $r=r_m$ the equation (32) reduces to the equation
for a harmonic oscillator with the energy $\tilde{E}=E-V(r_m)$,
and the energy spectrum is  \cite{me2002}
$$
E_n=\hbar \omega \biggl(n+\frac{1}{2}\biggr)
-\frac{GM(r_m)}{r_m}E_{Pl}
                                                                               \eqno(33)
$$
where $\omega^2=\Lambda c^2 \tilde{p}_{\perp}(r_m)$, and
$\tilde{p}_{\perp}$ is the dimensionless pressure normalized to
vacuum density $\rho_0$ at $r=0$; for the density profile (4)
$\tilde{p}_{\perp}(r_m)\simeq 0.2$.

The energy of zero-point vacuum mode \cite{me2002}
$$
\tilde{E}_0=\frac{\sqrt{3\tilde{p}_{\perp}}}{2}\frac{\hbar c}{r_0}
                                                                               \eqno(34)
$$
never exceeds the binding energy $V(r_m)$. It remarkably agrees
with the Hawking temperature from the de Sitter horizon
$kT_H=\frac{1}{2\pi}\frac{\hbar c}{r_0}$ \cite{GH}, representing
the energy of virtual particles which could become real in the
presence of the horizon. In the case of G-lump they are confined
by the binding energy $V(r_m)$ \cite{me2002}.

\subsection{Cosmological term as a source of mass}

The mass of both G-lump and $\Lambda$BH is directly connected to
cosmological term $\Lambda_{\mu\nu}$ by the ADM formula (9) which
in this case reads
$$
m=(2 G)^{-1} \int_0^{\infty}{\Lambda_t^t(r) r^2 dr}
                                                                              \eqno(35)
$$
and relates mass to the de Sitter vacuum at the origin
\cite{me2002}. In de Sitter-Schwarzschild geometry the parameter
$m$ is identified as a gravitational mass by flat asymptotic at
infinity. A mass is related to cosmological term, since de Sitter
vacuum appears as $r\rightarrow 0$.

This fact does not depend on
 extension of a cosmological term from $\Lambda g_{\mu\nu}$ to
$\Lambda_{\mu\nu}$. Whichever would be a matter source (2) for
this class of metrics and its interpretation (as associated with
$\Lambda_{\mu\nu}$ or not), mass is related to both de Sitter
vacuum trapped inside an object, and breaking of space-time
symmetry. De Sitter vacuum supplies an object with mass via smooth
breaking of space-time symmetry from the de Sitter group in its
center to the Lorentz group at its infinity \cite{me2002}.

This picture conforms with the basic idea of the Higgs mechanism
for generation of mass via spontaneous breaking of symmetry of a
scalar field vacuum. In both cases de Sitter vacuum is involved
and vacuum symmetry is broken. The gravitational potential $g(r)$
resembles a Higgs potential (see Fig.5).
\begin{figure}
\vspace{-8.0mm}
\begin{center}
\epsfig{file=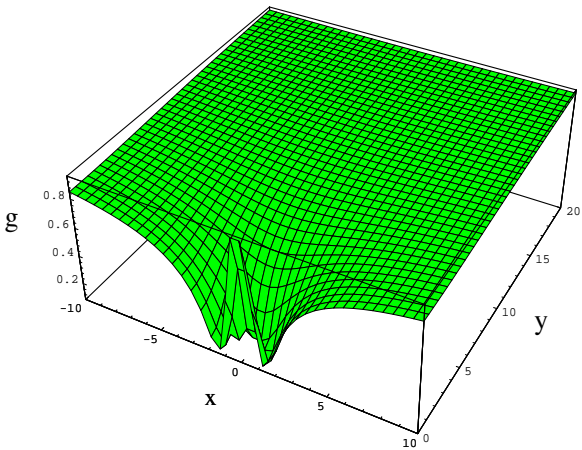,width=8.0cm,height=6.5cm}
\end{center}
\caption{The gravitational potential $g(r)$ for the case of
G-lump.}
\label{fig.5}
\end{figure}
Applying de Sitter-Schwarzschild geometry to estimate limits on
sizes of fundamental particles whose masses are related to de
Sitter vacuum through the Higgs mechanism, we get numbers  close
to experimental limits \cite{ethz}.
 Characteristic size in this geometry $r_s$ given by (25),
depends on vacuum density at $r=0$ and presents modification of
the Schwarzschild radius $r_g$ to the case when singularity is
replaced by de Sitter vacuum.  For the electron getting its mass
from the vacuum at the electroweak scale it gives
$r^e_s\sim{10^{-18}}$ cm, while the Schwarzschild radius is
$r^e_{g}\sim{10^{-57}}$ cm which is many orders of magnitude less
than $l_{Pl}$. In Fig.6 the geometrical lower limits on lepton
sizes estimated by (25), are compared with electromagnetic (EM)
and electroweak (EW) experimental upper limits\cite{ethz}. This
geometry gives also reasonable upper bound on the mass of the
Higgs scalar, $m_H\leq 154$ GeV \cite{ethz}.
\begin{figure}
\vspace{-8.0mm}
\begin{center}
\epsfig{file=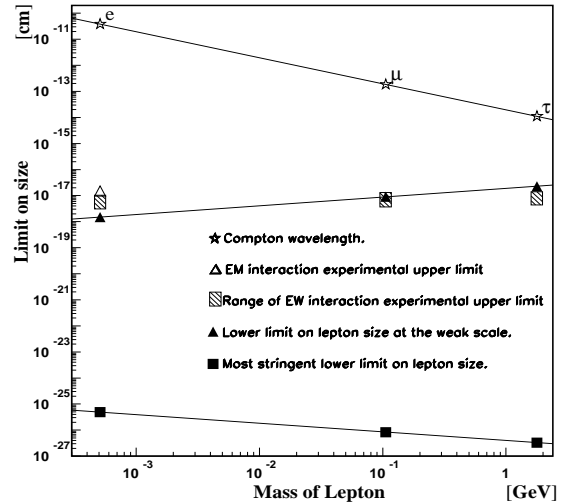,width=8.0cm,height=7.5cm}
\end{center}
\caption{Estimates of characteristic sizes for leptons
\protect\cite{ethz}. }
\label{fig.6}
\end{figure}

\section{Two-lambda geometry}

 \subsection{Types of configurations}

In the case of non-zero value of cosmological constant at
infinity, there are two vacuum scales, $\Lambda$ at the center and
$\lambda < \Lambda$ at infinity, and geometry has at most three
horizons \cite{us2002}. These are the internal Cauchy horizon
$r_{-}$, the black hole event horizon $r_{+}$, and the
cosmological horizon $r_{++}$. The number of horizons depends on
the mass parameter $m$ and on the parameter
$q=\sqrt{\Lambda/\lambda}$. The horizon-mass diagram is plotted in
Fig.7.
\begin{figure}
\vspace{-8.0mm}
\begin{center}
\epsfig{file=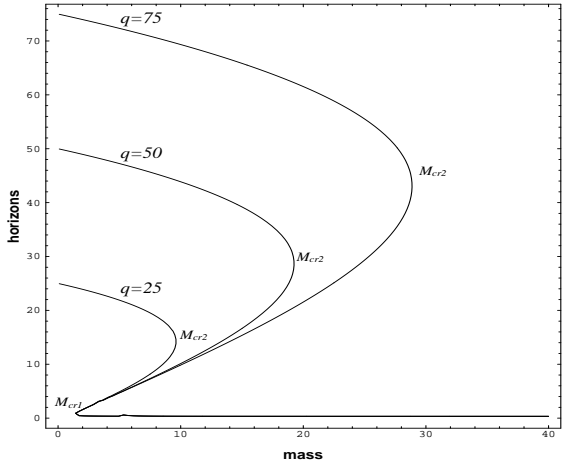,width=8.0cm,height=6.5cm}
\end{center}
\caption{Horizon-mass diagram of two-lambda geometry.}
\label{fig.7}
\end{figure}
In the range of masses $M_{cr1}<m<M_{cr2}$, geometry has three
horizons and describes a vacuum cosmological nonsingular black
hole. The global structure of space-time contains an infinite
sequence of $\Lambda$BH and $\Lambda$WH, their future and past
regular cores asymptotically de Sitter with $\Lambda g_{\mu\nu}$
as $r\rightarrow 0$, and asymptotically de Sitter universes with
$\lambda g_{\mu\nu}$ as $r\rightarrow \infty$ in the regions
$~{\cal CC}~$ between cosmological horizons and space-like
infinities \cite{us97}.

Three-horizon configuration represents the nonsingular
modification of the Kottler-Trefftz solution \cite{kot} known as
the Schwarzschild-de Sitter geometry \cite{GH}. The case
$m=M_{cr2} ~~  (r_{+}=r_{++})$ is the nonsingular modification of
the Nariai solution \cite{nariai}.

The case $m=M_{cr1} ~~  (r_{+}=r_{-})$ is another extreme black
hole state which appears due to replacing a singularity with a de
Sitter core. The critical value of mass $M_{cr1}$ at which
$r_{-}=r_{+}$, puts the lower limit on a black hole mass. It
practically does not depend on the parameter
$q=\sqrt{\Lambda/\lambda}$ and is given by (31).

 Five types of configurations
described by $\Lambda_{\mu\nu}$ geometry are shown in Fig.8
\cite{us97} for the case $q\equiv{\sqrt{{\Lambda}/{\lambda}}}=10$.
\begin{figure}
\vspace{-8.0mm}
\begin{center}
\epsfig{file=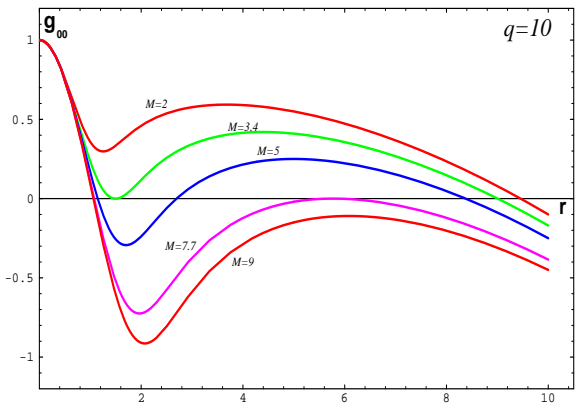,width=8.0cm,height=6.0cm}
\end{center}
\caption{Two-lambda configurations. The parameter $M$ is a mass
normalized to $(3/G^2 \Lambda )^{1/2}$. }
\label{fig.8}
\end{figure}
The case $M=2$ can be seen as a particle-like structure at the de
Sitter background $\lambda g_{\mu\nu}$. On the other hand, the
global structure in this case is the same as for de Sitter
geometry, so that in cosmological coordinates it represents
cosmological model of de Sitter type on global structure but with
cosmological density $\Lambda^t_t$ evolving from $\Lambda$ at the
beginning to $\lambda$ at late times. Similar case is for $M=9$.
Those two cases differ only on dynamics (nontrivial behavior in
the R region in the first case and in the T region in the second).

\subsection{$\Lambda_{\mu\nu}$ dominated cosmologies}

All cosmological models dominated by variable cosmological term
$\Lambda_{\mu\nu}$ belong to the Lemaitre class of cosmologies
with anisotropic perfect fluid described by
$$
ds^{2}=c^{2}d\tau^{2}-e^{\mu(R,\tau)}dR^{2}-r^{2}(R,\tau)d\Omega^{2}
                                                                            \eqno(36)
$$
and governed by equations \cite{us2001}
$$
\kappa p^{\Lambda}_r r^2=e^{-\mu}r'^{2}-2r\ddot{r}- \dot{r}^{2}-1
                                                                             \eqno(37a)
$$
$$
2\kappa p_{\perp}^{\Lambda} r= 2 e^{-\mu}r^{\prime\prime}-
e^{-\mu} r^{\prime}\mu ^{\prime} - \dot{\mu}\dot{r}-\ddot{\mu}r
                                                                           \eqno(37b)
$$
$$
\kappa \rho^{\Lambda} r^2=-e^{-\mu}
\left(2rr''+r'^{2}-rr'\mu'\right)+
\left(r\dot{r}\dot{\mu}+\dot{r}^{2}+1\right)
                                                                          \eqno(37c)
$$
$$
e^{\mu(R,\tau)}=\frac{{r^{\prime}}^2}{1+f(R)}
                                                                          \eqno(37d)
$$
The dot denotes differentiation with respect to $\tau$ and the
prime with respect to $R$.

The equation of motion describing the evolution \cite{us2001}
$$
{\dot r}^2+2r{\ddot r}+\kappa p_{r}^{\Lambda}  r^2=f(R)
                                                                          \eqno(38)
$$
It has the first integral
$$
\dot{r}^{2}=A + e^{\mu(R,\tau)}r(R,\tau)+f(R)r
                                                                           \eqno(39)
$$
and the second integral
$$
\tau-\tau_0(R)=\int_{r_0}^{r}{\sqrt{\frac{x}{A+e^{\mu(x)}x+f(R)}}}dx
                                                                             \eqno(40)
$$
Here $\tau_{0}(R)$ is an arbitrary function called the "bang-time
function". In the case of the Tolman-Bondi dust-filled model the
evolution is described by
$r(R,\tau)=(9GM(R)/2)^{1/3}(\tau-\tau_0(R))^{2/3}$, where
$\tau_0(R)$ represents the big bang singularity surface for which
$r(R,\tau)=0$.

 Near the regular surface $r=0$ corresponding to $R+\tau = -\infty$ the metric (36)
 takes the FRW form with the de Sitter
scale factor $a(\tau)\sim{\cosh(H_0 \tau)}$ for $f(R)<0$,
$a(\tau)\sim{\exp(H_0\tau)}$ for $f(R)=0$,
$a(\tau)\sim{\sinh(H_0\tau)}$ for $f(R)>0$, where $H_0$ is the
Hubble parameter corresponding to the initial value of $\Lambda$.
The evolution starts with the nonsingular non-simultaneous de
Sitter bang \cite{us2001,us2002}.  The inflationary start is
followed by a Kasner-type stage with contraction in the radial
direction and expansion in the tangential direction
\cite{us2001,us2002}
$$
ds^{2}=d\tau^{2}-F(R)(\tau+R)^{-2/3}dR^2-a(\tau+R)^{4/3}d\Omega^2
                                                                               \eqno(41)
$$
where $F(R)$ is a smooth regular function and $A$ is the constant
expressed in the model parameters. This is the stage when
acceleration of the "scale factor" $r(R,\tau)$ changes quickly and
drastically. In Fig.9 it is shown for the case of a spatially flat
model with $f(R)=0$ \cite{us2001}.
\begin{figure}
\vspace{-8.0mm}
\begin{center}
\epsfig{file=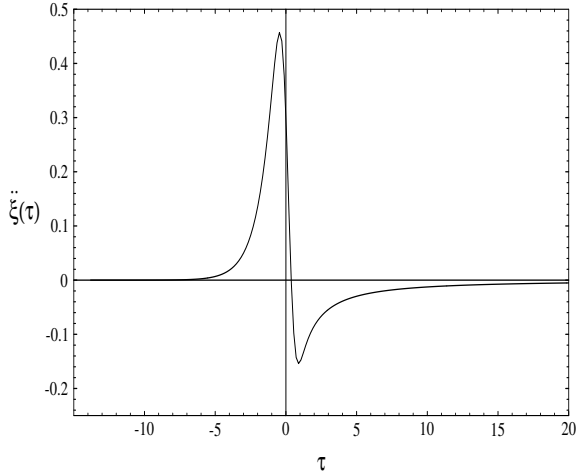,width=7.7cm,height=6.5cm,clip=}
\end{center}
\caption{The acceleration of the "scale factor" $r(\tau-\tau_0)$
normalized to $(GM/\Lambda)^{1/3}$.}
\label{fig.9}
\end{figure}
For a certain class of observers $\Lambda_{\mu\nu}$ dominated
models can be specified as Kantowski-Sachs models with regular R
regions.
 For the Kantowski-Sachs observers evolution starts from
horizons with a highly anisotropic "null bang" where the volume of
the spatial section vanishes. A null bung surface seems singular
to a comoving observer although it is perfectly regular in the
$\Lambda_{\mu\nu}$ geometry \cite{us2002}.

 In the case of the planar
spatial symmetry nonsingular $\Lambda_{\mu\nu}$ dominated
cosmologies are Bianchi type I, and  in the pseudospherical case
they are hyperbolic analogs of the Kantowski-Sachs models. At late
times all $\Lambda_{\mu\nu}$ dominated models approach de Sitter
asymptotic with $\lambda <\Lambda$ \cite{us2002}.

\section{Discussion}

$\Lambda_{\mu\nu}$ geometry follows from the Einstein spherically
symmetric equations with imposing the requirements of finiteness
of the ADM mass, regularity of density and dominant energy
condition on stress-energy tensor $T_{\mu\nu}$, identified, by its
symmetry, with the spherically symmetric vacuum associated with
$\Lambda_{\mu\nu}$. Dependently on the choice of coordinates and
range of parameters, $\Lambda_{\mu\nu}$ geometry describes vacuum
nonsingular black and white holes, particle-like vacuum
structures, and cosmologies with evolving vacuum density.

The inflationary cosmology specified various matter sources
associated with the Einstein cosmological term $\Lambda
g_{\mu\nu}$ (for review \cite{kolb}).

The question whether a regular black hole  can be obtained as a
false vacuum configuration described by $S=\int{d^4 x
\sqrt{-g}\biggl[R+(\partial \phi)^2-2V(\phi)\biggr]}$ with a
scalar field potential $V(\phi)$,  is addressed in the "no-go
theorem" \cite{dima}: Asymptotically flat regular black hole
solutions are absent for any non-negative potential $V(\phi)$.
This result has been generalized to the theory with the action
$S=\int{d^4 x\sqrt{-g}\biggl[R+F[(\partial \phi)^2,\phi]\biggr]}$,
where $F$ is an arbitrary function, to the multi-scalar theories
of sigma-model type, and to scalar-tensor and curvature-nonlinear
gravity theories \cite{kirill}. It has been shown that the only
possible regular solutions are either de Sitter-like with a single
cosmological horizon  or those without horizons, including
asymptotically flat ones. The latter do not exist for $V(\phi)\geq
0$, so that the set of causal false vacuum structures is the same
as known for $\phi=const$ case, namely Minkowski (or anti-de
Sitter), Schwarzschild, de Sitter, and Schwarzschild-de Sitter
\cite{kirill}, and thus does not include de Sitter-Schwarzschild
configurations.

In the case of {\it complex} massive scalar field the regular
structures can be obtained in the minimally coupled theory with
positive $V(\phi)$ \cite{Schunck}. The best example is boson stars
(\cite{Mielke} and references therein), but in this case algebraic
structure of the stress-energy tensor does not satisfy Eq.(21),
and asymptotic  at $r=0$ is not de Sitter.

The stress-energy tensor of structure (2) gives the regular
magnetic monopole solution in nonlinear electrodynamics
\cite{kirmm}. This is the first  candidate for a matter source
associated with the variable cosmological term $\Lambda_{\mu\nu}$.

\section* {Acknowledgement}

This work was supported by the Polish Committee for Scientific
Research through the grant 5P03D.007.20 and through the grant for
UWM.

\end{document}